\documentclass[aps,
twocolumn,
showpacs,prl
]{revtex4}
\usepackage{epsf}
\begin{document}
\title{Substrate Specificity of Peptide Adsorption: A Model Study}
\author{Michael Bachmann}
\email[E-mail: ]{Michael.Bachmann@itp.uni-leipzig.de}
\author{Wolfhard Janke}
\email[E-mail: ]{Wolfhard.Janke@itp.uni-leipzig.de}
\homepage[\\ Homepage: ]{http://www.physik.uni-leipzig.de/CQT.html}
\affiliation{Institut f\"ur Theoretische Physik, Universit\"at Leipzig,
Augustusplatz 10/11, D-04109 Leipzig, Germany}
\begin{abstract}
Applying the contact density chain-growth algorithm to lattice
heteropolymers, we identify the conformational transitions of 
a nongrafted hydrophobic--polar heteropolymer with 103 residues in the vicinity
of a polar, a hydrophobic, and a uniformly attractive substrate. Introducing
only two system parameters, the numbers of surface contacts and intrinsic hydrophobic contacts,
respectively, we obtain surprisingly complex temperature and solvent dependent, substrate-specific 
pseudo-phase diagrams.  
\end{abstract}
\maketitle
From recent experiments of the adsorption
of short peptides at semiconductor substrates it is known that different surface
properties (materials such as Si or GaAs, crystal orientation, etc.) as well as
different amino acid sequences strongly influence the binding properties of these peptides
at the substrate~\cite{whaley1,goede1}. This specificity will be of particular
importance for future
sensory devices and pattern recognition~\cite{bogner1} at the nanometer scale. The 
reasons for this binding specificity are far from being clear, and it is a big
challenge from the experimental and theoretical point of view to understand
the basic principles of substrate--peptide cooperativity. This problem can be seen as
embedded into a class of similar studies, where the adsorption and docking behavior
of polymers is essential, e.g., protein--ligand 
binding~\cite{ligand1}, prewetting and layering transitions in polymer solutions 
as well as dewetting of polymer films~\cite{wetting1}, molecular pattern, 
electrophoretic polymer deposition and growth~\cite{foo1}.    

The experimental
equipment has reached such a high resolution allowing for precise identification
of single molecule shapes at the substrate, and the available computational capacities
combined with sophisticated algorithms will 
make it possible to examine the problem of a hybrid interface between biological 
and inorganic materials~\cite{willett1} step by step. 

In a first step, we have recently analyzed the adsorption of a finite, nongrafted 
homopolymer at an attractive substrate in a cavity and discussed in detail the 
phase diagram in the thermodynamic limit, as well as
pseudo-transitions that depend strongly on the given number of monomers~\cite{bj1}.
Similar studies of grafted polymers are reported in Refs.~\cite{vrb1,singh1,prell1}.
For heteropolymers the thermodynamic limit is unreachable because of the
sequence of different types of monomers. This ``disorder-inducing'' sequence
renders the heteropolymer adsorption a distinguishingly different problem
to homopolymer substrate-binding.
The pseudo-phase transitions of the heteropolymer
system will strongly depend on three main points: the sequence,
the monomer-specific interaction with the substrate, and the total number of monomers
in the chain. In this work, we particularly focus on the substrate-specificity.
In order to reduce the complexity of the problem to a minimum, we study 
the hydrophobic-polar (HP) model~\cite{dill1}, where the heteropolymer
consists of a given sequence of only two types of monomers: hydrophobic
(H) and polar (P). 
We use the simplest form of the model, where only the hydrophobic force acts
and the number of nearest-neighbor contacts between H monomers being nonadjacent
along the chain, $n_{\rm HH}$, is related to the energy of the heteropolymer.
The interaction with the substrate is modeled in a like manner: The energy of the
heteropolymer is reduced by the number of nearest-neighbor contacts between the substrate
and those monomers that experience the attractive force of the substrate. For all
other monomers the influence of the substrate is only entropic. 

In order
to study the specificity of surface--binding, we investigate three attractive substrate 
models. In the first variant, all monomers, independent of their hydrophobic or polar 
character, are equally attracted by the substrate and the energy of the system is proportional
to the total number of monomer--surface contacts, $n_s^{\rm H+P}$. In the second
and third model, the substrate is either hydrophobic or polar, i.e., in the first
case only the hydrophobic monomers in the heteropolymer sequence are attracted,
and in the latter the attraction between substrate and heteropolymer dipoles
dominates. In these models, the respective hydrophobic substrate contacts, $n_s^{\rm H}$,
and polar surface contacts, $n_s^{\rm P}$, are energetically favored. Thus, the
models can be expressed in the energetic form
\begin{equation}
\label{model}
E_s(n_s,n_{\rm HH})=-\varepsilon_s n_s-\varepsilon_{\rm HH} n_{\rm HH},
\end{equation}     
where, depending on the substrate model, $n_s=n_s^{\rm H+P}$, $n_s^{\rm H}$, or 
$n_s^{\rm P}$. For our qualitative study, it is sufficient to choose for all three
models the same energy scales $\varepsilon_s = 1$ and $\varepsilon_{\rm HH}=s$, 
where $s$ denotes the solubility which controls the solvent quality (the larger
the value of $s$, the worse the solvent). Since we are interested in the fluctuations
of the respective contact numbers with respect to temperature $T$ and solubility $s$,
we define the contact density as 
$g(n_s,n_{\rm HH}) = \delta_{n_s 0}\,g^{\rm u}(n_{\rm HH})+
(1-\delta_{n_s 0})g^{\rm b}(n_s, n_{\rm HH})$,
with the contributions of the densities of conformations without ($g^{\rm u}$) and with
($g^{\rm b}$) contact to the substrate. In order to regularize the influence of the
unbound conformations and for computational efficiency, the heteropolymer is restricted
to reside in a cage, i.e., in addition to the physically interesting 
attractive surface there is a steric, neutral
wall parallel to it in a distance $z_w$. The value of $z_w$ is chosen sufficiently large 
to keep the influence on the unbound heteropolymer small (in this work we used $z_w=200$). 
Introducing the partition
sum $Z_{T,s}\sim \sum_{n_s,n_{\rm HH}} g(n_s,n_{\rm HH})\exp(-E_s/k_BT)$ and denoting thermodynamic
expectation values of a quantity $O(n_s,n_{\rm HH})$ by $\langle O\rangle$,
the contact correlation matrix $M_{xy}(T,s)=\langle xy\rangle_c=\langle xy\rangle-\langle x\rangle\langle y\rangle$
with $x,y=n_{\rm HH},n_s$ separates the fluctuations of the
surface and hydrophobic contacts according to the respective energy scale vector 
$(\varepsilon_{\rm HH},\varepsilon_s)=(s,1)$, and therefore the energetic fluctuations are
accounted for in the specific heat, defined by (from now on we set $k_B\equiv 1$)
\begin{equation}
\label{corrmat}
C_V(T,s)=\frac{1}{T^2}(s,1)M(T,s)\left(\begin{array}{c}s\\ 1\end{array}\right).
\end{equation}
The heteropolymer sequence we chose in our study possesses 103 monomers (37 being hydrophobic,
66 polar) as introduced in Ref.~\cite{103lat} and often used for benchmark tests of new 
algorithms~\cite{103lat,103toma,hsu1,bj3}. The advantage is that this heteropolymer forms
a nicely compact hydrophobic core in the low-energy conformations (as depicted for $s=1$ and 
$n_s=0$, e.g., in Ref.~\cite{bj3}), completely screened from the solvent by a shell of polar monomers. 
\begin{figure}
\centerline{\epsfxsize=8.3cm \epsfbox{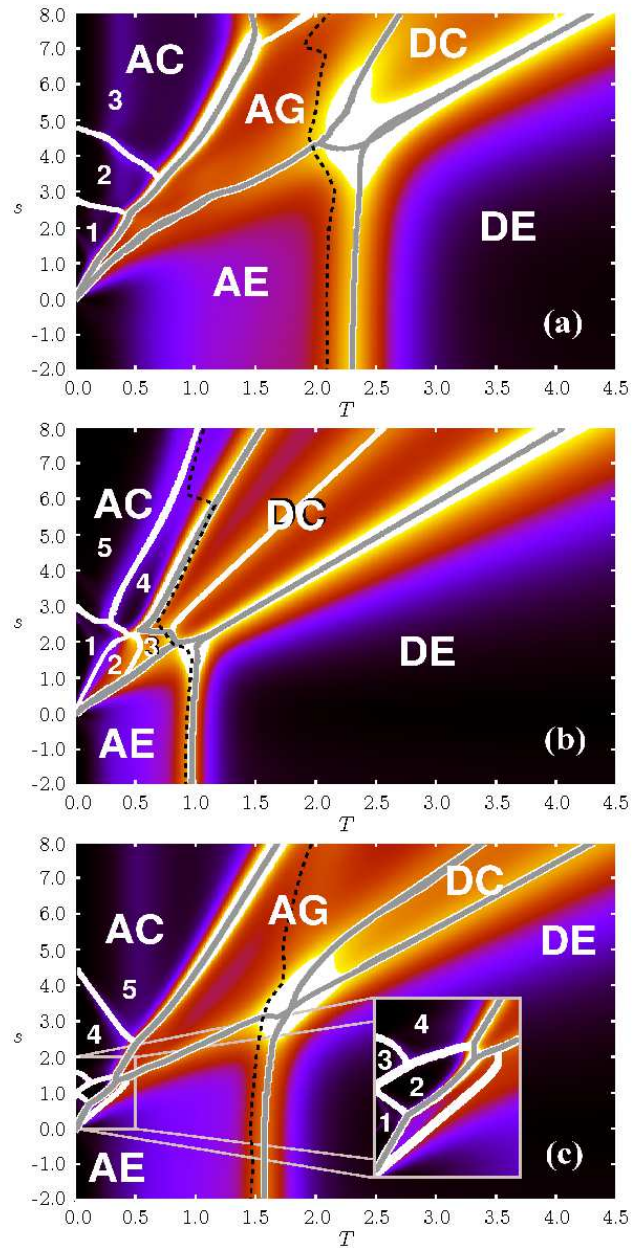}}
\caption{\label{pd103} (Color online) Pseudo-phase diagrams of the 103mer
near three different substrates that are attractive for 
(a) all, (b) only hydrophobic, and (c) only polar monomers.
}
\end{figure}

In order to calculate the contact densities for the three systems,
we have applied an enhanced version of the multicanonical chain-growth algorithm~\cite{bj4,prell2}.
In contrast to move-set based Metropolis Monte Carlo or conventional chain-growth 
methods which would require many separate simulations to obtain 
results for different parameter pairs $(T,s)$ and which frequently suffer from slowing down 
in the low-temperature sector, our method allows the computation of the {\em complete}
contact density for each system within a {\em single} simulation run. Since the contact
density is independent of temperature and solubility, energetic quantites
such as the specific heat can easily be calculated for all values of $T$ and $s$ (nonenergetic
quantities require accumulated densities to be measured within the simulation, but this is also 
no problem). 

In Figs.~\ref{pd103}(a)--(c) the contour profiles of the specific heats for the different
substrates are shown (the brighter the color the larger the value of $C_V$). We interpret the ridges (for accentuation marked
by white and gray lines) as the boundaries of the pseudo-phases. It should be noted, however, 
that in such a finite system the exact positions of active regions exhibited by fluctuations of other quantities 
usually deviate, but the qualitative behavior is similar~\cite{bj3}. The gray lines indicate the
main transition lines, while the white lines separate pseudo-phases that strongly depend on specific properties 
of the heteropolymer, such as its exact number and sequence of hydrophobic and polar monomers. As a first result,
we have found that the binding-unbinding transition appears to be first-order like. Assuming the 
contact numbers $n_s$ and $n_{\rm HH}$ to be kind of order parameters adequately describing the 
state of the heteropolymer, we define the contact free energy as 
$F_{T,s}(n_s,n_{\rm HH}) = -T\ln\, \left[g(n_s,n_{\rm HH})\exp(-E_s/T)\right]$ and 
the probability for a macro-state with $n_s$ substrate and $n_{\rm HH}$ hydrophobic contacts as
$p_{T,s}(n_s,n_{\rm HH})=g(n_s,n_{\rm HH})\exp(-E_s/T)/Z_{T,s}$.
Close to the binding-unbinding transition,
adsorbed and desorbed states coexist. This is exhibited by two clearly separated minima of the
contact free energy $F_{T,s}(n_s,n_{\rm HH})$. In the figures we have marked the coexistence line, where
both minima take the same value, by the dashed black lines. At lower temperatures, the most probable
conformation is an adsorbed one, while for higher temperatures desorbed conformations dominate. 

Despite the surprisingly rich and complex phase behavior there are main ``phases'' that can be
distinguished in all three systems. 
These are separated in Figs.~\ref{pd103}(a)--(c) by gray lines.
Comparing the three systems we find that they all possess
pseudo-phases, where adsorbed compact (AC), adsorbed expanded (AE), desorbed compact (DC), and
desorbed expanded (DE) conformations dominate, similar to the generic phase diagram of the 
homopolymer~\cite{bj1,vrb1,singh1,prell1}. ``Compact'' here 
means that the heteropolymer has formed a 
dense hydrophobic core, while expanded conformations exhibit dissolved, random-coil like structures.
The sequence and substrate specificity of
heteropolymers generates, of course, a rich set of new interesting and selective phenomena
not available for homopolymers. One example is the pseudo-phase of adsorbed globules (AG), which is noticeably present
only in those systems, where all monomers are equally attractive to the substrate (Fig.~\ref{pd103}(a)) and
where polar monomers favor contact with the surface (Fig.~\ref{pd103}(b)). In this phase, the conformations
are intermediates in the binding-unbinding region. This means that monomers 
currently desorbed from the substrate have not yet found their position within a compact conformation.
Therefore, the hydrophobic core, which is smaller than in the respective adsorbed phase 
(i.e., at constant solubility $s$), appears as a loose cluster of hydrophobic monomers. 
\begin{figure}
\centerline{\epsfxsize=8.8cm \epsfbox{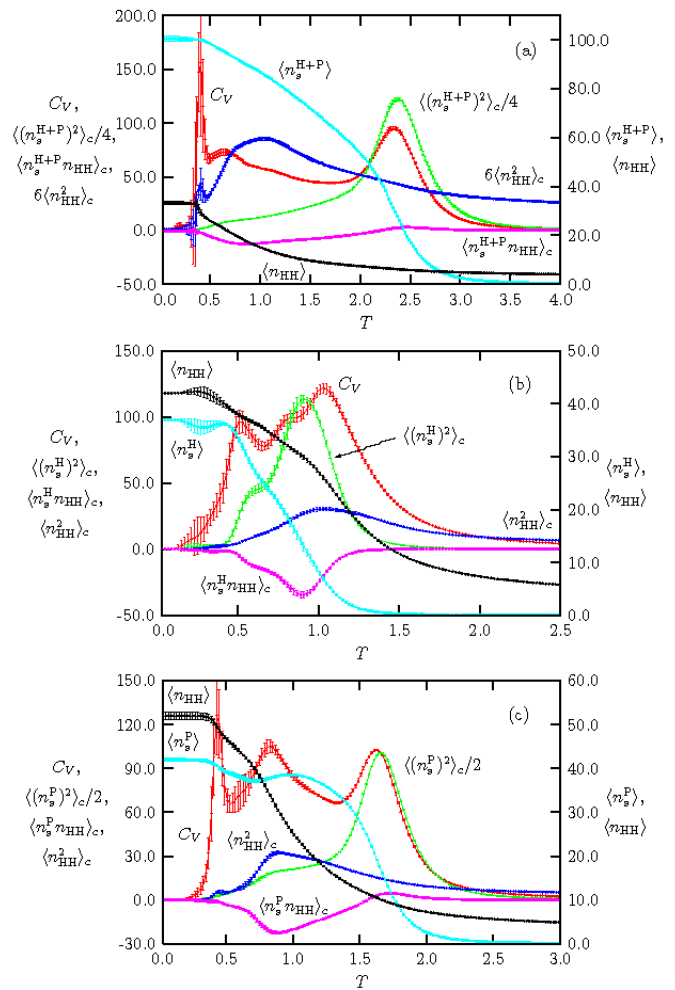}}
\caption{\label{c103} (Color online) 
Temperature dependence of specific heat, correlation matrix components, and 
contact number expectation values of the 103mer for surfaces attractive for 
(a) all, (b) only hydrophobic, and (c) only polar monomers at $s=2$. 
}
\end{figure}

In Figs.~\ref{c103}(a)--(c), we have plotted, exemplified for $s=2$,  
the statistical averages of the contact numbers $n_s$ and $n_{\rm HH}$
as well as their self- and cross-correlations $M$ for the three systems. For comparison 
we have also included the specific heat, 
whose peaks correspond to the intersected transition lines of
Figs.~\ref{pd103}(a)--(c) at $s=2$. From Figs.~\ref{c103}(a) and (c) we read off that the
transition from AC to AG near $T\approx 0.4$ is mediated by fluctuations of the 
intrinsic hydrophobic contacts. The very dense hydrophobic domains in the AC subphases
lose their compactness. This transition is absent in the hydrophobic-substrate system
(Fig.~\ref{c103}(b)). The signal seen belongs to a hydrophobic layering AC subphase transition,
which influences mainly the number of surface contacts $n_s^{\rm H}$.
The second peak of the specific heats belongs to the transition between adsorbed compact or
globular (AC, AG) and expanded (AE) conformations. This behavior is similar in all
three systems. Remarkably, it is accompanied by a strong anti-correlation between 
surface and intrinsic contact numbers, $n_s$ and $n_{\rm HH}$. Not surprisingly,
the hydrophobic contact number $n_{\rm HH}$ fluctuates stronger than the number
of surface contacts, but apparently in a different way. Dense conformations with
hydrophobic core (and therefore many hydrophobic contacts) possess a relatively small
number of surface contacts. Vice versa, conformations with many surface contacts
cannot form compact hydrophobic domains. Finally, the third specific-heat peak 
marks the binding-unbinding transition, which is, as expected, due to a strong
fluctuation of the surface contact number.

The strongest difference between the three systems is their behavior in pseudo-phase AC, which is
roughly parameterized by $s>5T$.
If hydrophobic and polar monomers are equally attracted by the substrate (Fig.~\ref{pd103}(a)),
we find three AC subphases in the parameter space plotted. In subphase AC1, film-like conformations
dominate, i.e., all 103 monomers are in contact with the substrate. Due to the good solvent quality in 
this region, the formation of a hydrophobic core is less attractive than the maximal deposition of all monomers 
at the surface, the ground state being $(n_s^{\rm H+P},n_{\rm HH})_{\rm min}=(103,32)$. In fact, instead of
a single compact hydrophobic core there are nonconnected hydrophobic clusters. At least on the used
simple cubic lattice and the chosen sequence, the formation of a single hydrophobic core is necessarily
accompanied by an unbinding of certain polar monomers and, in consequence, an extension of the
conformation into the third spatial dimension. In fact, this happens when entering AC2 
[$(n_s^{\rm H+P},n_{\rm HH})_{\rm min}=(64,47)$], 
where a single hydrophobic
two-layer domain has formed at the expense of losing surface contacts. In AC3,
the heteropolymer has maximized the number of hydrophobic contacts and only local arrangements of monomers 
on the surface of the very compact structure lead to the still possible maximum number of substrate
contacts. $F_{T,s}$ is minimal for $(n_s^{\rm H+P},n_{\rm HH})_{\rm min}=(40,52)$.

The behavior of the heteropolymer adsorbed at a surface that is only attractive to
hydrophobic monomers (Fig.~\ref{pd103}(b)) is apparently different in the AC phase. 
Since surface contacts 
of polar monomers are energetically not favored, the subphase structure is determined
by the concurrence of two hydrophobic forces: substrate attraction and formation of intrinsic
contacts. In AC1, the number of 
hydrophobic substrate contacts is maximal for the single 
hydrophobic layer, $(n_s^{\rm HH},n_{\rm HH})_{\rm min}=(37,42)$. The {\em single} two-dimensional hydrophobic domain is also
maximally compact, at the expense of displacing polar monomers into a second layer.
In subphase AC2, intrinsic contacts are entropically 
broken with minimal free energy for $35\le n_{\rm HH}\le 40$, while $n_s^{\rm HH}=37$
remains maximal. 
Another AC subphase, AC3, exhibits a
hydrophobic layering transition at the expense of hydrophobic substrate contacts. Much more interesting
is the subphase transition from AC1 to AC5. The number of hydrophobic substrate 
contacts $n_s^{\rm HH}$ of the ground-state conformation dramatically decreases 
(from $37$ to $4$) and the hydrophobic monomers collapse
in a one-step process from the compact two-dimensional domain to the maximally compact
three-dimensional hydrophobic core. The conformations are mushroom-like structures grafted 
at the substrate. AC4 is similar to AC5, with advancing desorption.

Not less exciting is the subphase structure of the heteropolymer interacting with a polar substrate
(Fig.~\ref{pd103}(c)).
For small values of $s$ and $T$, the behavior of the heteropolymer is dominated by the concurrence
between polar monomers contacting the substrate and hydrophobic monomers favoring the formation
of a hydrophobic core, which, however, also requires cooperativity of the polar monomers.
In AC1, film-like conformations ($n_s^{\rm P}=66$, $n_{\rm HH}=31$)
with disconnected hydrophobic clusters dominate. Entering AC2, hydrophobic contacts 
are energetically favored and a second hydrophobic layer forms at the expense
of a reduction of polar substrate contacts 
[$(n_s^{\rm P},n_{\rm HH})_{\rm min}=(61,37)$]. 
In AC3, the upper layer 
is mainly hydrophobic [$(n_s^{\rm P},n_{\rm HH})_{\rm min}=(53,45)$], 
while the poor quality of the solvent ($s$ large) and the comparatively 
strong hydrophobic force let the conformation further collapse [AC4: $(n_s^{\rm P},n_{\rm HH})_{\rm min}=(42,52)$]
and the steric cooperativity forces more 
polar monomers to break the contact to the surface and to form a shell surrounding the hydrophobic
core [AC5: $(n_s^{\rm P},n_{\rm HH})_{\rm min}=(33,54)$]. 
  
Summarizing, we have performed a detailed analysis of the pseudo-phase diagrams in the $T$--$s$ plane for
a selected heteropolymer with 103 monomers in 
cavities with an adsorbing substrate being either attractive independently of the monomer type,
or selective to hydrophobic or polar monomers, respectively. Beside the expected
adsorbed and desorbed phases, we find a rich subphase structure in
the adsorbed phases with compact hydrophobic domains, which is specific to heteropolymers. In particular, the formation 
of layered subphases in the low-temperature region depends mainly
on the quality of the solvent.

Here, we have mainly focused on the contact numbers $n_s$ and $n_{\rm HH}$,
but the study of structural quantities, such as the gyration tensor, which exhibits
a phase-dependent asymmetry in the components parallel and perpendicular to the substrate, 
confirms our interpretation of the subphase-behavior of the system~\cite{bj5}. 
Since current experimental equipment is capable to reveal molecular structures at 
the nanometer scale, it should be possible to investigate the grafted structures
dependent on the solvent quality. 
This is essential for answering the question under what circumstances 
binding forces are strong enough to refold peptides or proteins. The vision of future 
biotechnological and medical applications is fascinating as it ranges from protein-specific 
sensory devices to molecular electronic devices at the nanoscale.     

We thank A.\ Beck-Sickinger, K.\ Goede, and M.\ Grundmann for interesting discussions.
This work is partially supported by the DFG (German Science Foundation) grant  
under contract No.\ JA 483/24-1. Some simulations were performed on the 
supercomputer JUMP of the John von Neumann Institute for Computing (NIC), Forschungszentrum
J\"ulich, under grant No.\ hlz11. 
\end{document}